\documentclass[english]{paper}
\usepackage[T1]{fontenc}
\usepackage[latin9]{inputenc}
\usepackage{float}
\usepackage{graphicx}

\makeatletter
\usepackage{chngpage}

\makeatother

\usepackage{babel}

\begin{document}

\title{Targeting by Transnational Terrorist Groups}

\author{Alexander Gutfraind}
\maketitle
\begin{abstract}
Many successful terrorist groups operate across international borders
where different countries host different stages of terrorist operations.
Often the recruits for the group come from one country or countries,
while the targets of the operations are in another. Stopping such
attacks is difficult because intervention in any region or route might
merely shift the terrorists elsewhere. Here we propose a model of
transnational terrorism based on the theory of activity networks.
The model represents attacks on different countries as paths in a
network. The group is assumed to prefer paths of lowest cost (or risk)
and maximal yield from attacks. The parameters of the model are computed
for the Islamist-Salafi terrorist movement based on open source data
and then used for estimation of risks of future attacks. The central
finding is that the USA has an enduring appeal as a target, due to
lack of other nations of matching geopolitical weight or openness.
It is also shown that countries in Africa and Asia that have been
overlooked as terrorist bases may become highly significant threats
in the future. The model quantifies the dilemmas facing countries
in the effort to cut such networks, and points to a limitation of
deterrence against transnational terrorists.
Address: Center for Nonlinear Studies and T-5, Theoretical Division,
Los Alamos National Laboratory, Los Alamos, New Mexico 87545 USA.
agutfraind.research@gmail.com
\end{abstract}

\section{Introduction}

Despite vast investments in counter-terrorism, victory in the global
war on terror remains elusive. In part this is because terrorist groups
are highly adaptive in their tactics and strategy. When airport scanners
were installed to detect weapons and explosives, terrorists switched
to explosives that cannot be detected using the scanners and to other
modes of attack \cite{Landes78,Enders04}. When it became harder to
reach US soil or attack US embassies groups shifted to attacks against
other countries or less fortified installations \cite{Enders05,Woo04}.
Like international businesses, globalized terrorist groups are vast
international enterprises that tap into the most successful business
practices and cost-efficient solutions \cite{Arquilla01}. If a country
erects high barriers to entry or develops an effective domestic counter-terrorism
response then terrorists switch their targeting to a safer and more
accessible place. If a country no longer provides a haven for recruitment,
training and planning of operations, those will shift elsewhere\cite{Hoffman06}.

\begin{figure}[H]
\begin{centering}
\includegraphics[bb=0bp 0bp 720bp 424bp,clip,width=0.6\paperwidth]{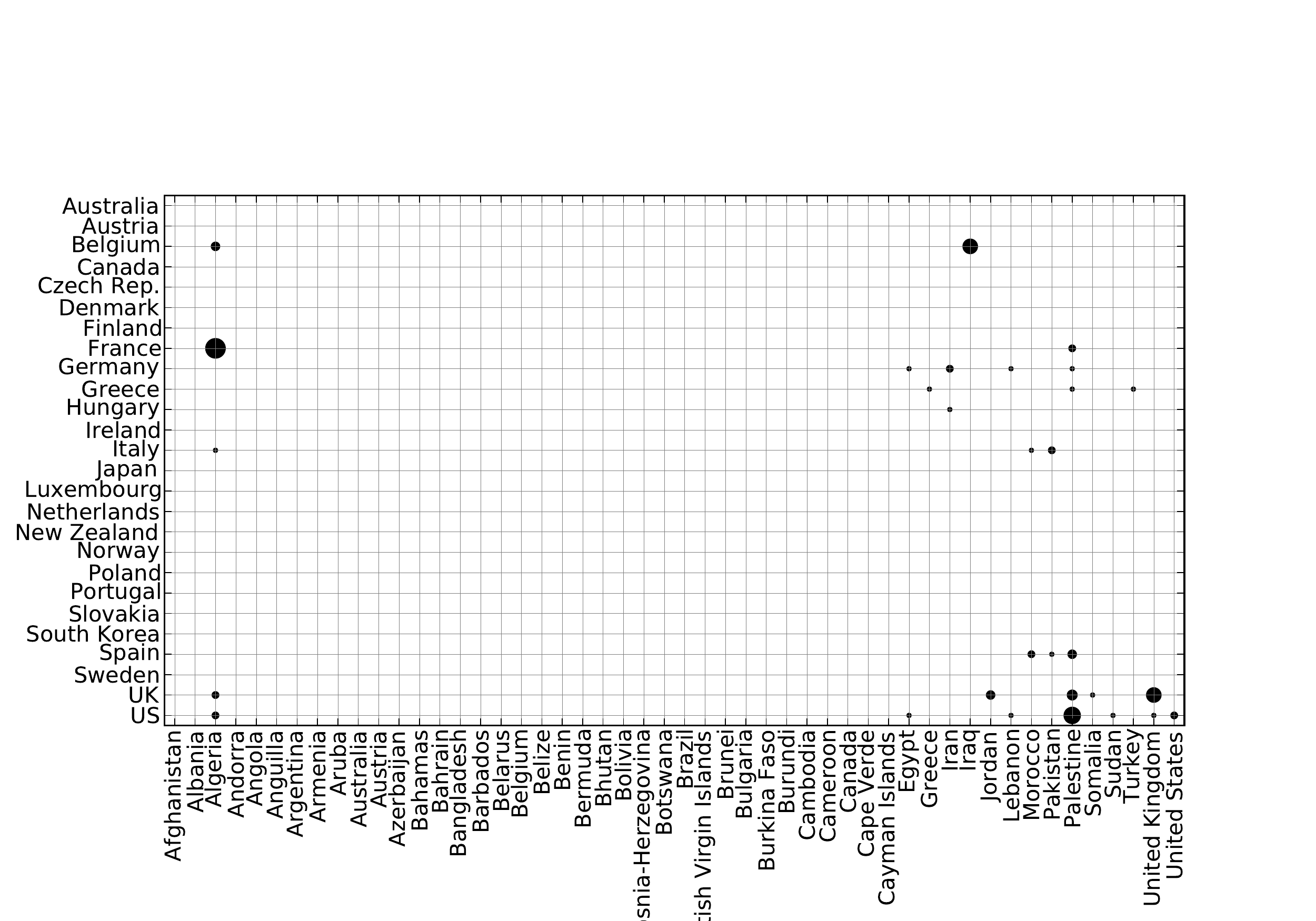}
\par\end{centering}

\caption{Terrorist attacks by Islamist groups over 1990-2008. The fraction
of all transnational plots that originated at country $i$ and targeted
country $j$ is proportional to the area of a circle at coordinates
($i,j$). The source countries on the horizontal axis account for
$>99\%$ of all attacks against OECD countries (vertical).\label{fig:ITERATE}}

\end{figure}

Adaptability makes risk estimation challenging. One possible basis
for risk assessment is extrapolation of historical data, such as the
ITERATE dataset of transnational attacks \cite{Micholus09iterate}.
Fig.~\ref{fig:ITERATE} shows all ITERATE attacks carried out by
Islamist groups on OECD countries in which the national origins of
the attackers and the target country are known. Many of the incidents
in the matrix are due ethno-nationalist conflicts, such the GIA attacks
in France or due to attacks by home-grown cells inspired by Salafis.
While a substantial fraction of attacks were against the US, many
attacks also targeted France, the UK and other countries. 

In the rest of the paper we will introduce another method for risk
assessment: a quantitative network-based model. The model takes demographic
and economic information pertaining to violent Islamist-Salafi groups
- the most probable source of future attacks - and estimates the risk
of various transnational terror plots. The model suggests that the
future of transnational terrorism may be substantially different from
the past:
\begin{enumerate}
\item several regions will become large new sources of transnational terrorism
(sec.~\ref{sec:Predictions}),
\item the US will rise as the terrorists' preferred attack destination (sec.~\ref{sec:Predictions}),
\item successes in stopping foreign-based plots against the US will increase
the threat to other countries (subsec.~\ref{sub:National-Fortresses}),
and
\item deterrence will be hard to achieve (subsec.~\ref{sub:Deterrence}).
\end{enumerate}
The model is based on an activity network for stages of terrorist
attacks. The network represents decisions required for terrorist operations
on the global scale, such as which country to attack. No distinction
is made between {}``transnational'' and {}``international'' terrorism,
both referring to terrorist groups that operate using foreign bases,
support or inspiration. This coarsened scale of analysis exposes the
strategic picture and can guide counter-terrorism decision making
at the national and international levels. It also quantifies a kind
of unintended effect from counter-terrorism measures known as {}``transboundary
externality'' \cite{Sandler03,Sander04}: the redirection of terrorists
from one country to another, because the latter is less protected.

To the author's knowledge, this is the first model of its kind in
the open literature. Previous work considered target selection by
terrorists but where targets are implicitly within a single country
so the costs of bringing the attackers and their weapons to their
target are negligible (see e.g. \cite{Bier07}). In contrast, for
transnational terrorism security measures and international logistics
play a central role in attack planning \cite[Ch.3]{Harmon00}. Other
work considered the structure of the terrorist networks at the level
of individual operatives or functions, rather than as a global network
(cf. \cite{Hartnell89,Lindelauf08covert,Corman06,Gutfraind09cascades,Woo09}.)

It is sometimes argued that such economic models of terrorism are
unreliable because terrorism is an irrational behavior by fanatics.
However, the preponderance of evidence supports the alternative view
- the rational choice theory (RCT) \cite{Shugart09}. RCT claims that
terrorist groups and leaders are rational agents capable of strategic
decision-making. Their decisions are expressions of {}``instrumental
rationality'', that is, consistent with their values and objectives
\cite{Lake02}. The sophistication and technological adaptability
of terrorists, such as in developing triggers for explosives, is strong
evidence for their intelligence \cite{Hoffman06,Hanson08}. More evidence
for RCT comes from studies of target selection \cite{Sander04}. Those
consistently find evidence for a substitution effect - as governments
improve protection to certain targets, terrorists substitute them
with less protected targets \cite{Anderton05,Dugan05,Hanson08}. Indeed,
the defining feature of terrorism - the use of violence against civilians
rather than against military targets - is a strategic substitution
effect because the latter are harder targets. Another line of evidence
for rationality comes from analyzing the internal dynamics of terrorist
groups. Rather like non-violent organizations, they perform cost-benefit
analyses, employ financial controls and run financial audits \cite{Shapiro08,Shapiro09}.

\section{A Model of Transnational Terrorism}

We now describe a model of a transnational terrorist groups. Such
groups are characterized by their global aims, as opposed to regional
conflicts; they recruit and attack throughout the world. Islamic fundamentalist
groups will serve as the central application of this model. These
include al-Qaida, Hezbollah but also possibly extremist groups not
currently thought to be violent, such as Hizb-ut-Tahrir. Rather than
considering specific groups and limiting to their current bases of
operation, this study will consider every region of the world that
currently or may in the future host violent Islamic groups. We focus
on those groups because they are probably the most potent present-day
transnational terrorist threat. The model could also evaluate other
violent transnational ideologies if we re-estimate its parameters.

\subsection{Operations Submodel}

Suppose a transnational group controls a cell in a country, and must
decide where to dispatch this cell (the cell might also be self-mobilizing,
in which case it must solve its own targeting problem.) The three
options are (1) do a domestic attack, (2) send the cell to attack
another country, and (3) do nothing. Option (1) entails certain risks
and costs for collecting intelligence and preparing weapons. Dispatching
the cell into another country, (2), incurs the additional cost and
risk of interception due to security barriers, such a visas, intelligence
collection in a foreign environment, and cultural difficulties. However,
the other country might have more favorable security environment or
offer better targets - more significant or less protected. Option
(3) - abandon the attack and hide - has little or no risk or accounting
cost and preserves the cell for future operations.

\begin{figure}
\begin{centering}
\includegraphics[width=0.6\paperwidth]{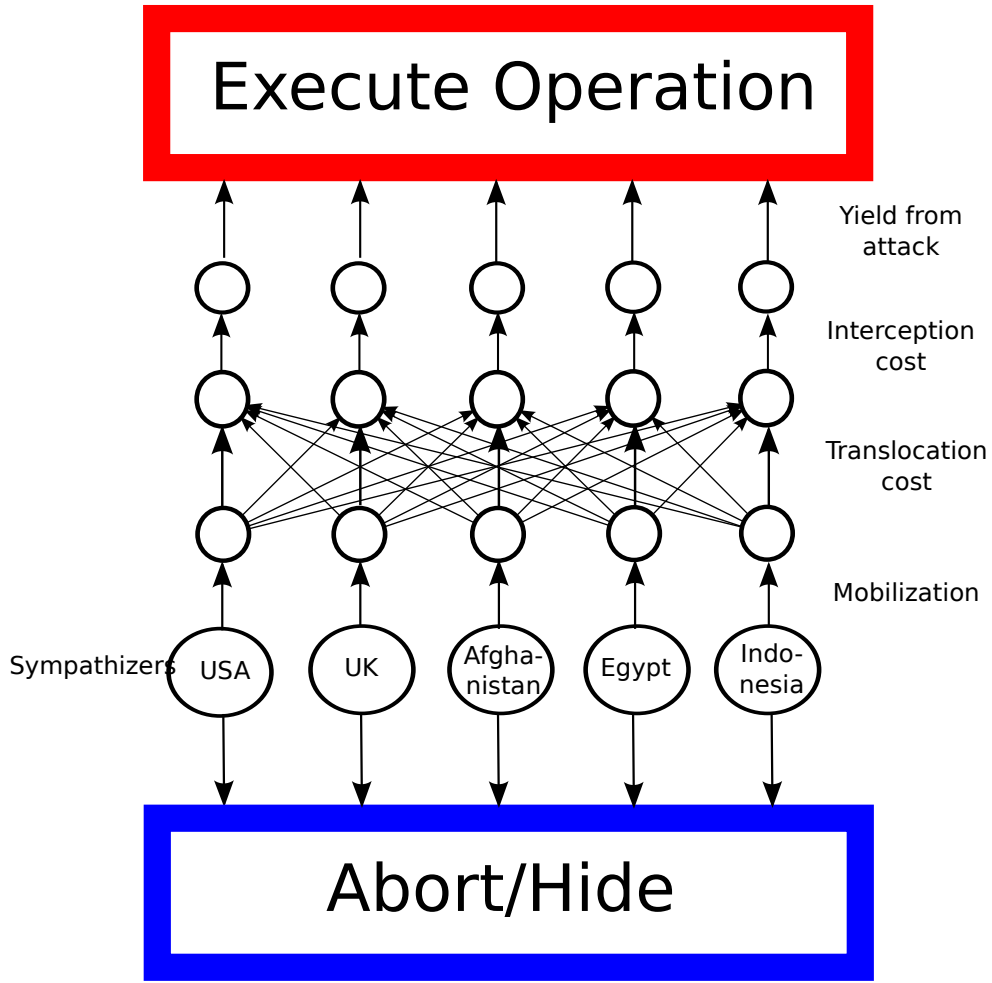}
\par\end{centering}

\caption{Illustration of the model for $5$ countries. The vertical direction
represents countries while the horizontal represents different stages
in execution of plots. Domestic attack plans correspond to motion
upwards, while transnational attacks also make a diagonal transition.
The full model includes many more countries\label{fig:model-illus}.}

\end{figure}

Any rational decision maker must weight the costs and benefits and
take the action offering the greatest net benefit. Surely then terrorists
would also do such analysis, weighing at least the most obvious target
choices and travel routes. A simple way of representing this is with
an activity network, where nodes represent different stages of terrorist
operations at different countries, and edges show the cost and risk
involved in each stage (Fig.~\ref{fig:model-illus}). In Fig.~\ref{fig:model-illus},
the columns correspond to the countries, including the country of
origin, while the rows correspond to the postures of the cells: the
stages of the plots.

The network represents the options of the rational decision maker
as directed paths - chains of nodes and directed edges that start
in the source country node and lead to either the {}``attack'' node
or the {}``abandon/hide'' node. If complete information is available
about the costs and benefits of each option, then the rational decision
is to select the path with the highest utility, that is the path with
highest net benefit (benefit minus cost.) For any path $p$, the cost
$c(p)$ is found by adding the weights on the edges (tasks) constituting
the path. 

The edge weights of this network could represent resources like money
or materiel that are consumed and produced by terrorist operations.
The network could also be used to perform a probabilistic risk assessment
(PRA): to evaluate the gains from possible operations and the probabilities
of successfully completing intermediate stages in the operations.
Such a PRA is what we will do. In other words we will take the perspective
of the terrorists: determine what they want and what they fear in
order to anticipate how they will act%
\footnote{Here is how PRA is represented by networks. Suppose in a multi-stage
terrorist operation $r_{s}$ is the probability of success at stage
$s$ (out of $k$ stages in total) conditional on success at every
previous stage. Suppose the gain from a successful operation is $G$
($\geq1$). Then the \emph{expected} gain from the operation is $E=r_{1}r_{2}\dots r_{k}G$.
Let us now relate $r_{s}$ values to costs ($c_{s}\geq0$) using exponentiation:
$r_{s}=e^{-c_{s}}$, and let the gain be a function of yield $Y$:
$G=e^{-Y}$. Thus, an attack has expected gain $E=\exp\left[-\left(c_{1}+c_{2}+\dots+c_{k}+Y\right)\right]$.
In the network representation of terrorist operations, we can compute
the sums in the exponent by adding edge weights along network paths
that trace through all the stages. Paths of lower weights translate
to attacks of greater expected gain. By comparing such paths we could
anticipate which attacks would have the highest expected gain. %
}. 

It is an open question whether terrorist groups can or will use such
an algebraic method to analyze their operations. However, given their
sophistication they may well come to the same decisions using other
means or through operational experience. Of course they might also
intentionally avoid the most probable attacks to achieve surprise,
but only at a cost (and models could also be constructed to anticipate
that too.)

Consider now the following specific model for transnational terrorist
attacks, Fig.~\ref{fig:model-illus}. A cell that was mobilized at
country $i$ experiences (1) the translocation cost/risk, $T_{ij}$,
representing the barriers for moving from country $i$ to country
$j$; (2) the risk of interception at country $j$, $I_{j}$; and
(3) the yield $Y_{j}$ from attacks at country $j$. Yield reflects
the gain to the terrorists from a successful attack, and so has the
opposite sign from cost. A domestic attack at country $i$ has cost
$c(p)=T_{ii}+I_{i}+Y_{i}$ while a transnational attack has cost $c(p)=T_{ij}+I_{j}+Y_{j}$
($i\neq j$). Because negative edge weights are costs in the sum that
represent risks, the words {}``cost'' and {}``risk'' will be used
interchangeably. Sometimes attackers reach country $j$ through one
or several intermediate countries (exploiting e.g. the Schengen treaty),
a possibility we ignore for simplicity. From the counter-terrorism
point of view, the likelihood of a particular plot depends also on
the supply of plots originating at each country. Therefore, we will
estimate for each country $i$ the number of cells that originate
there, $S_{i}$.  If the group decides to abandon, its path has cost
$c(p)=A$. The value of $A$ may be a negative, representing the preservation
of the cell, or positive, if the cell cannot be reactivated. 

It is possible to include additional costs like cost of recruitment
or training in the model but this will be left for future studies
because the data is hard to estimate. Here it was assumed that the
costs are negligible and indeed many cells involved in recent attacks
recruited themselves and trained with information obtained from websites.
Another possible extension is attacking country $j$ through its foreign
representatives: embassies, officials and even tourists. Such attacks
represent a different type of attack path and could be easily added,
albeit by adding more hard-to-estimate parameters to the model. Similarly,
one may consider attack on modes of international transit such as
airplanes and ships.

The model's parameters can be estimated from open source information
with a modest degree of confidence (see section \ref{sec:Estimation-of-Parameters}).
Briefly, transit costs were estimated from data on global migration,
the risk of interception from national expenditure on internal security
and attack yields based on the political power of the targeted country,
represented by its GDP. The supply of plots is estimated from public
opinion surveys measuring support for terrorist attacks and demographic
data.

\subsection{Stochastic Decisions Submodel}

If transnational terrorist groups could determine the values of the
parameters precisely (the next section discusses this problem), then
they should be able to plot the optimal attack from each country $i$
by considering all possible options and finding the path that minimizes
cost: \[
\min_{j}\left[\underbrace{T_{ij}+B_{j}+Y_{j}}_{c(p_{ij})}\right]\,.\]

However, one of the general difficulties in decision making is uncertainties
about costs and risks. Terrorists, like other decision makers should
therefore occasionally identify the optimal attack incorrectly. Reliable
risk assessment must therefore take into account the possibility that
adversaries make mistakes (as well as use unpredictability to achieve
surprise - but that is even harder to assess.) Fortunately, suitable
stochastic prediction methods have already been developed for activity
network models like in Fig.~\ref{fig:model-illus}. With those methods
probabilities can be assigned to different terrorist plans based on
the costs of the corresponding paths on the network. We use the model
in \cite[Ch.3]{Gutfraind09thesis} based on Markov chains. In it the
path of least cost is typically assigned the highest probability but
other paths have non-zero probabilities, and these probabilities can
be quite high (for details see Appendix, sec.~\ref{sub:Computation-of-Probabilities}).%
\footnote{One of the advantages of the stochastic model is that it can interpolate
between the two extremes of complete ignorance and perfect information
using a single parameter $\lambda$ ($\ge0$) that describes the amount
of information available to the adversaries. For a given level of
information, the probability that a path $p$ would be selected is
proportional to $\exp(-\lambda\, c(p))$. When $\lambda$ is very
large the path of least cost has a much higher probability than any
of the alternatives, while $\lambda$ close to $0$ assigns all paths
approximately the same probability. We set $\lambda=0.1$ in the following
but its value has a smooth effect on the predicted plots (i.e. the
sensitivity is low). A value of $0.1$ means that if the terrorist
group learns of a increase in path cost by $10$ units, its probability
of taking the path will decrease by a multiplicative factor of $\approx2.72$.
The effect depends on the original path probability: it is not as
great a decrease when the original probability is high.%
}

From this Markov chain model it is possible to compute the number
of times any particular country (represented by its attack node) would
be targeted as well as to compute the changes in targeting due to
various defensive actions, which are represented as increases in edge
weights. It is also possible to determine whether defensive actions
would materially increase the costs for the adversaries or merely
lead them to change targets.

\section{Estimation of Parameters\label{sec:Estimation-of-Parameters}}

The model contains several sets of inputs: (1) the supply of plots
at country $i$, $S_{i}$; (2) barriers for moving from country $i$
to country $j$ $T_{ij}$; (3) risk of interception at country $j$,
$I_{j}$; and (4) the yield from attacks at country $j$, $Y_{j}$.
The yield of abandoning, $A$ will be set to $\infty$ (no abandoned
plots) and its effect will be analyzed separately. Because (1)-(4)
contain security-related information that is also difficult to measure,
the information is not published. Luckily, one can derive estimates
from publicly-available demographic and economic data. Readers wishing
to see the final results of estimates should skip to tables \ref{tab:country-tables},
\ref{tab:country-supply} and \ref{tab:Transit-cost} of the Appendix
and ignore the rest of this section. 

In building the estimates, it will be assumed for simplicity that
each stage of terrorist operations carries about the same amount of
risk. Namely, that the medians of $T_{ij}$ ($i\neq j$) and of $I_{j}$
both equal$1$. Of course, some plots will be much less risky than
others because both $T_{ij}$ and $I_{j}$ have considerable variability.
The yield from attacks is also normalized relative to the median.
Transformations of this kind on costs and yields are unavoidable if
we wish to remove the effect of units, but they do reduce the reliability
of the model. However, the core findings of the model regarding certain
countries agree well with intuition, as will be seen.

\subsection{Estimating the Supply of Plots, $S_{i}$}

To assess the threat from the greater violent Islamic movement it
is not sufficient to consider the support for a particular group like
al-Qaida. Even if we assumed that al-Qaida has no support in certain
countries, other groups might mobilize supporters there. Another concern
is self-mobilization: modern terrorist groups sometimes avoid active
recruitment and instead provide inspiration and guidance while relying
on self-radicalization to provide them with foot soldiers \cite{Sageman08}.

A comprehensive picture on possible plots can be obtained from surveys.
Over the last decade most recently in 2009 the Pew charitable trusts
run several global attitude surveys which among others, asked Muslims
about their support for suicide bombings \cite{Pew09radical}. In
each of the surveyed countries respondents were asked to state whether
suicide bombings is {}``never justified'' ($\sigma^{n}$), {}``rarely
justified'' ($\sigma^{r}$), {}``sometimes justified'' ($\sigma^{s}$),
and {}``often justified'' ($\sigma^{o}$). These are given as fractions
of the respondents. For some countries no data was available, so the
quantities were extrapolated from countries in the same geographic
region (e.g. Middle East, Americas etc.) Pew also collected data on
the Muslim population in 235 different countries and territories ($J_{i}$)
\cite{Pew09muslim} (of course the overwhelming majority of Muslim
everywhere are opposed to terrorism in the name of their religion).
The supply of violent plots can be estimated by taking the population
and multiplying by the weighted fraction of respondents professing
support for violence (the weights are $s_{r},s_{s},s_{o}$). One must
also take into account that only a small fraction of those who profess
radical ideology would actually be involved in a plot and that several
people are involved in each plot (a factor $Q$): \[
S_{i}:=Q\cdot J_{i}\cdot\left(s_{r}\sigma^{r}+s_{s}\sigma^{s}+s_{o}\sigma^{o}\right)\,.\]

The support weights were set by default based on the assumption that
every increase in professed support leads to an increase by a factor
of $2$ in the resources available to terrorist groups: ($s_{r},s_{s},s_{o}$)=($0.25,0.50,1.00$).
Appendix~\ref{sec:Parameter-Estimates} explores the sensitivity
of supply $S_{i}$ to this assumption through two alternatives: most
of the tangible support come from the narrow but committed minority,
($s_{r},s_{s},s_{o}$)=($0.1,0.2,1.0$), and a situation where even
the least-committed supporters materially boost the terrorists, ($s_{r},s_{s},s_{o}$)=($0.33,0.66,1.00$).
Notes that the weights effect only the relative importance of regions
as sources of terrorism, not the targets of plots originating in a
given region.

The factor $Q$ is dependent on social and tactical issues, and hence
should not vary much across countries. Because $Q$ enters as a multiplicative
term at all source countries, its value has no bearing on the relative
risk estimates. Nevertheless, it could be crudely estimated as follows.
In 2006 the head of the British Security Service (MI5) reported that:
{}``... my officers and the police are working to contend with some
200 groupings or networks, totaling over 1600 identified individuals
(and there will be many we don't know) who are actively engaged in
plotting, or facilitating, terrorist acts here and overseas'' \cite{Manningham-Buller06}.
Furthermore {}``over 100,000 of our citizens consider that the July
2005 attacks in London were justified.'' This implies active participation
at a rate of at least $1.6\%$ and $8$ people per plot ($Q=0.002$).

\subsection{Estimating the Barriers for Moving from Country $i$ to Country $j$,
$T_{ij}$}

Barriers to transnational attacks include both deliberate barriers
such as screening and intelligence, and unofficial barriers such as
differences in language and culture. Official barriers depend on factors
such as the intelligence available on targets in the destination country,
the cooperation the targeted country received from both the country
of departure and the transport agent (e.g. airline). None of those
figures are publicly available but a proxy measure can be found, as
follows. Transnational terrorists often use tourism, education or
immigration as cover to obtain travel documents and permits. Indeed,
travel in all of those categories became more difficult across the
developed world as a result of the security measures introduced after
the 9/11 attacks. Migration patterns thus provide an estimate of official
barriers. Unofficial barriers to migration are likewise similar to
the unofficial barriers to terrorism, including differences in language,
culture, ethnicity and others. Therefore, the foreign-born migrant
population, suitably normalized, could be used as a proxy of transnational
freedom of travel. Migration into most OECD countries is documented
by the OECD \cite{OECD06migration}.\\
It is to be expected that the number of migrants would be positively
correlated with the population of the countries and negatively correlated
with distance. This is known as a {}``gravity law'' model. Many
national and international relationships such as trade flows are well-approximated
by gravity laws \cite{Erlanger90,Jung08}, named for their similarity
to Newton's law for the force of gravity. Therefore an estimate for
the number of migrants between country $i$ and country $j$ is the
product of their populations (data: UN) divided by their distance
squared (data: CEPII \cite{mayer06}): $\frac{p_{i}p_{j}}{d_{ij}^{2}}$.
When the actual number of migrants, $m_{ij}$, falls below this estimate,
that may indicate heightened official or unofficial barriers. Thus
we define the raw transnational terrorism barrier between countries
$i$ and $j$ as: \[
\widehat{T_{ij}}:=\frac{p_{i}p_{j}}{d_{ij}^{2}}/m_{ij}\,.\]
The data must now be standardized, for several reasons: (1) the barrier
data should be comparable with other costs considered by the terrorists
(and in the model) by removing the effects of units for population
and distance; and (2) since the barrier is a cost and risk, it must
be represented by positive number in the activity network. Therefore,
the quantity $T_{ij}$ was computed from $\widehat{T_{ij}}$ by determining
the minimum ($Min\widehat{T}$) and median ($Med\widehat{T}$). Because
of (2) $MinT$ is subtracted from $\widehat{T_{ij}}$, and for (1)
the quantity is divided by the median of the shifted values:\[
T_{ij}=\left(\widehat{T_{ij}}-Min\widehat{T}\right)/\left(Med\widehat{T}-Min\widehat{T}\right).\]
The resulting values have a median of $1.0$. The more standard procedure
of first removing the average and dividing by the standard deviation
was rejected because the distribution is visibly non-Gaussian with
large positive outliers (hence a skewed mean and large variance).
Domestic operations will be assumed to have negligible barriers ($T_{ii}=0$
for all countries $i$). The OECD data lacks information about migration
to non-OECD countries. Therefore set $T_{ij}=\infty$ in all such
cases (effectively blocking such paths). 

While the OECD includes some of the most geopolitically-important
nations of the planet, obtaining data on translocation costs to non-OECD
countries would be valuable for two reasons. First, only with such
data can we estimate the risk of terrorism to those nations (such
as the July 11 bombings in Kampala), and second to estimate the effect
of counter-terrorism policies in the OECD on terrorism in other countries.
Indeed, in the ITERATE database of terrorist incidents \cite{Micholus09iterate},
attacks on OECD countries that can be traced to Islamist groups account
for only 22\% of all Islamist attacks.

\subsection{Estimating the Risk of Interception at Country $j$, $I_{j}$}

The risk of interception can be estimated from OECD data on expenditure
on public order and safety as percentage of GDP. The relevant figure
is the fraction of GDP rather than the raw figure because the number
of valuable targets is related to the size of the economy, so the
fractional figure indicates the 

level of security vulnerable sites can receive. The GDP also correlates
with the population size (in rich countries) and thus to the amount
of police resources available per person. The extreme case of totalitarian
police states is suggestive: in such countries internal security expenditures
are disproportionately large relative to the GDP, and indeed terrorists
have a lot of difficulties operating there \cite{Harrison09}. This
estimate of course neglects the efficiency of internal security force
- a factor that is hard to estimate. 

The internal security data is transformed almost exactly like the
barrier data and for the same reasons: start with figures for internal
security expenditure as a fraction of GDP for country $j$, $SEC_{j}$
then compute the minimum ($MinSEC=\min_{j}SEC_{j}$) and median ($MedSEC$),
and then normalize:\[
I_{j}=\left(SEC_{j}-MinSEC\right)/\left(MedSEC-MinSEC\right)\,.\]

Generally speaking, we find that there is not much variation in the
risk of interception in different countries (raw data is in the Appendix,
sec.~\ref{sec:Parameter-Estimates}), as compared to variation in
factors such as yield, discussed next. This suggests that interception
risk plays a minor role in target choice.

\subsection{Estimating the Yield from Attacks at Country $j$, $Y_{j}$ }

Transnational terrorist attacks attempt to influence policies. For
example, one of al-Qaida's original objectives was to compel the withdrawal
of US forces from Arabia, while Hezbollah forced France and the US
to withdraw their peacekeepers from Lebanon in the 1980s. The precise
value of targets shifts with time and the political situation, but
typically larger richer countries make for more powerful players in
the international arena, and hence more important targets. Moreover,
their homelands carry more targets of symbolic, political and economic
significance. The economic damage from the loss of life and physical
assets is also higher in richer countries because they tend to have
higher productivity for labor and capital. Thus, it is expected that
transnational terrorists would seek to attack larger richer countries.
The weight of a country can be estimated from its dollar GDP figures
at current exchange rates (source: UN data).

Timing or political dynamics does play a role in transnational terrorism
but its importance might be overestimated. For example, the Madrid
2004/03/11 train bombings are often viewed as intending to pressure
the Spanish government to withdraw its forces from Iraq, and they
were timed with the Spanish elections. But surely an important factor
was Spain's geopolitical weight (GDP is ranked 12th in the world)
and its large contribution to the 2003 invasion. Otherwise al-Qaida
could have just as well pressured smaller countries such as El Salvador
and Mongolia to withdraw their contributions to the invasion. 

Here is how the yield $Y_{j}$ was computed from the GDP figures.
Recall that costs (barriers, internal security) are all positive,
so yields must be negative. Let the minimum GDP be $MinGDP$, and
the median $MedGDP$. The following formula produces negative values
with a standardized median:\[
Y_{j}=\left(MinGDP-GDP_{j}\right)/\left(MedGDP-MinGDP\right)\,.\]
The resulting values have a median of $-1.0$.

\section{Predictions\label{sec:Predictions}}

\begin{figure}[!tbph]
\begin{centering}
\includegraphics[bb=0bp 0bp 720bp 424bp,clip,width=0.6\paperwidth]{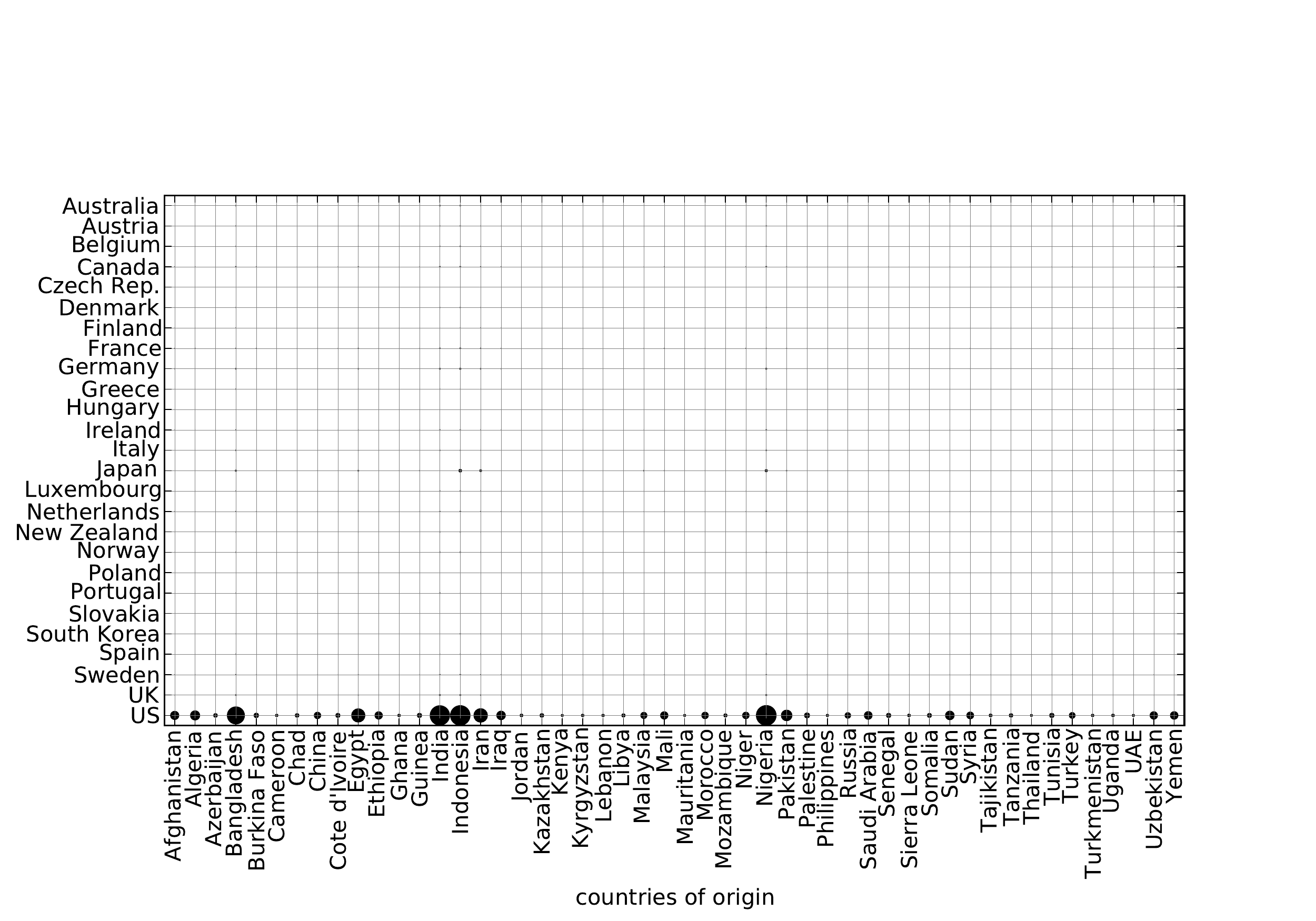}
\par\end{centering}

\caption{Predicted attack risk matrix. The area of a circle at coordinates
($i,j$) is proportional to the number of plots from $i$ to $j$.
The bottom row indicates that the vast majority of plots target the
US.\label{fig:baseline} }

\end{figure}

One way of representing the solution of the model is through an attack
matrix, the counterpart of the historical data matrix in Fig.~\ref{fig:ITERATE}.)
The model predicts (Fig.~\ref{fig:baseline}) that the United States
would attract the bulk of transnational terrorism - all other countries
are almost free of terrorism (white squares). The reason the United
States is such a magnet is because of its vast geopolitical weight
and relatively open borders. 

Examination of the sources for attacks exposes a number of risks.
There is a considerable terrorist threat from Bangladesh, India, Indonesia
and Nigeria. They combine a large population and relatively high support
for terrorism. It is notable that the 2009 Christmas bomber was Nigerian
- one of the first attacks on OECD from that country. 

The high burden of attacks borne by the US is directly related to
the rational choice model: if there is a big prize to be won by attacking
the US, no rational terrorist would attack other countries. The reasons
why non-US attacks do occur (cf. Fig.~\ref{fig:ITERATE}) include:
(1) some terrorist groups such as ethno-nationalist groups see as
their enemy a particular country and lack a global strategy; (2) global
Salafi groups have not yet expanded their recruitment channels in
countries such as Nigeria and Bangladesh, so a large fraction of attacks
is still carried out by groups with more narrow agendas; and (3) the
US has deployed counter-terrorism measures commensurate with the threat
it faces, making the US too costly to attack.

One implication of this finding concerns US policy. The matrix justifies
in principle outlays by the US government towards countering international
terrorism as a whole, without regard to its target. Investments in
international counter-terrorism measures, such as policing, if effective
in reducing the number of plots, are also efficient from the US perspective
because the US, being the target of choice, will retain most of the
benefits from reducing the terror threat \cite{Lee88}. Unfortunately,
many policies previously adopted were ineffective or had perverse
effects on international terrorism (the so-called {}``blowback'')
(see e.g. \cite{Ganor05,Ganor08}.)

\subsection{National Fortresses\label{sub:National-Fortresses}}

\begin{figure}[H]
\begin{centering}
\includegraphics[bb=0bp 0bp 660bp 424bp,clip,angle=270,origin=c,width=0.6\paperwidth]{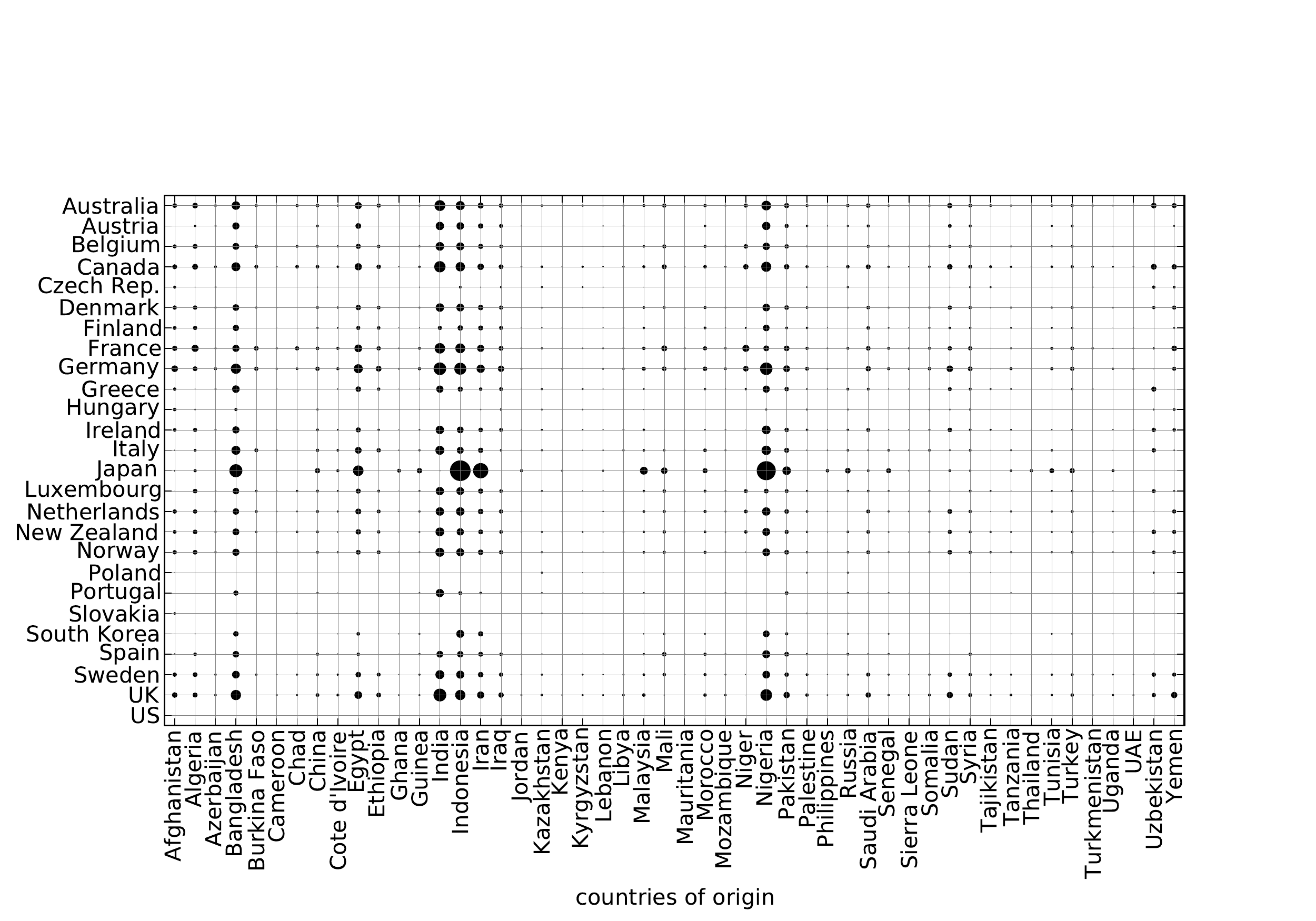}
\par\end{centering}

\caption{Attack risk matrix in scenario where US becomes inaccessible to foreign
plots. Terrorist plots increase in all other OECD countries. \label{fig:fortress}}

\end{figure}

Consider now several alternative scenarios for the future, motivated
by strengthening of counter-terrorism defenses, which may make transnational
attacks less feasible. Suppose the US was successful in deterring
attacks against itself by greatly increasing the barriers to entering
US soil. If so, Fig.~\ref{fig:fortress} shows the likely effect. 

The protection of US frontiers will significantly\emph{ }increase
the attack risk to most other OECD nations because transnational groups
should then switch to more accessible targets. Perhaps surprisingly,
Japan, now rarely mentioned as a target will see the largest absolute
increase in terrorism. This prediction is due to its international
profile, Japan being the second largest country in the OECD on several
measures. Japan's woes will be shared to some extent by most other
major OECD countries, who will also see an increase in attacks.

\begin{figure}[!tbph]
\begin{centering}
\includegraphics[bb=0bp 0bp 720bp 424bp,clip,width=0.6\paperwidth]{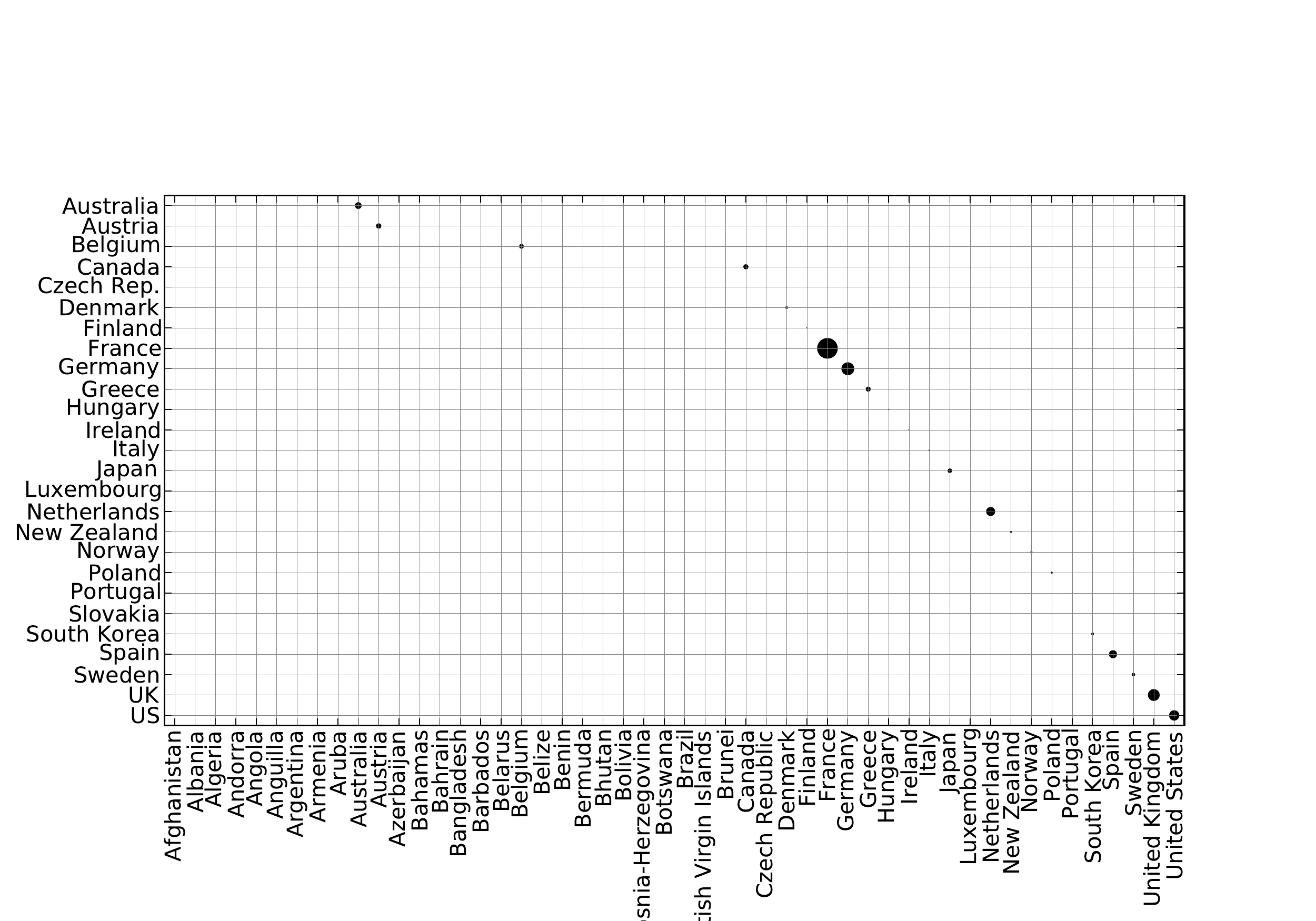}
\par\end{centering}

\caption{Attack matrix in a scenario where OECD countries cannot be accessed
by foreigners (both inside and outside the OECD). Countries with relatively
large radicalized Muslim populations (e.g. France) rise in rank relative
to their OECD peers. The total number of attacks on OECD countries
must decrease significantly because foreign plots are blocked.\label{fig:home-grown}}

\end{figure}

Another possible scenario is where the security forces in each country
are able to intercept the majority of external plots against their
homelands. In other words, the translocation cost becomes very large
($T_{ij}=\infty$ for $i\neq j$). In this world, the dominant form
of terrorism is home-grown. As Fig.~\ref{fig:home-grown} shows,
this materially changes the risk matrix. Countries such as the France,
with relatively large and relatively radicalized Muslim communities
will see much more terrorism.

The two scenarios point to large conflicts of interest between OECD
countries in tackling transnational terrorism. Helping the US intercept
plots through advance warning will increase terrorism everywhere else.
More broadly, country A will not always benefit from helping country
B. Doing so might sometimes increase the chances that A's enemies,
some of which even based in B, will shift to B. This factor may explain
part of the difficulty achieving intelligence sharing and international
police cooperation. Indeed B could even come to an understanding with
its home-grown terrorists in which they abstain from domestic attacks
in return for non-intervention in their activities.

\subsection{Deterrence\label{sub:Deterrence}}

\begin{figure}[!h]
\begin{centering}
\includegraphics[width=0.6\paperwidth]{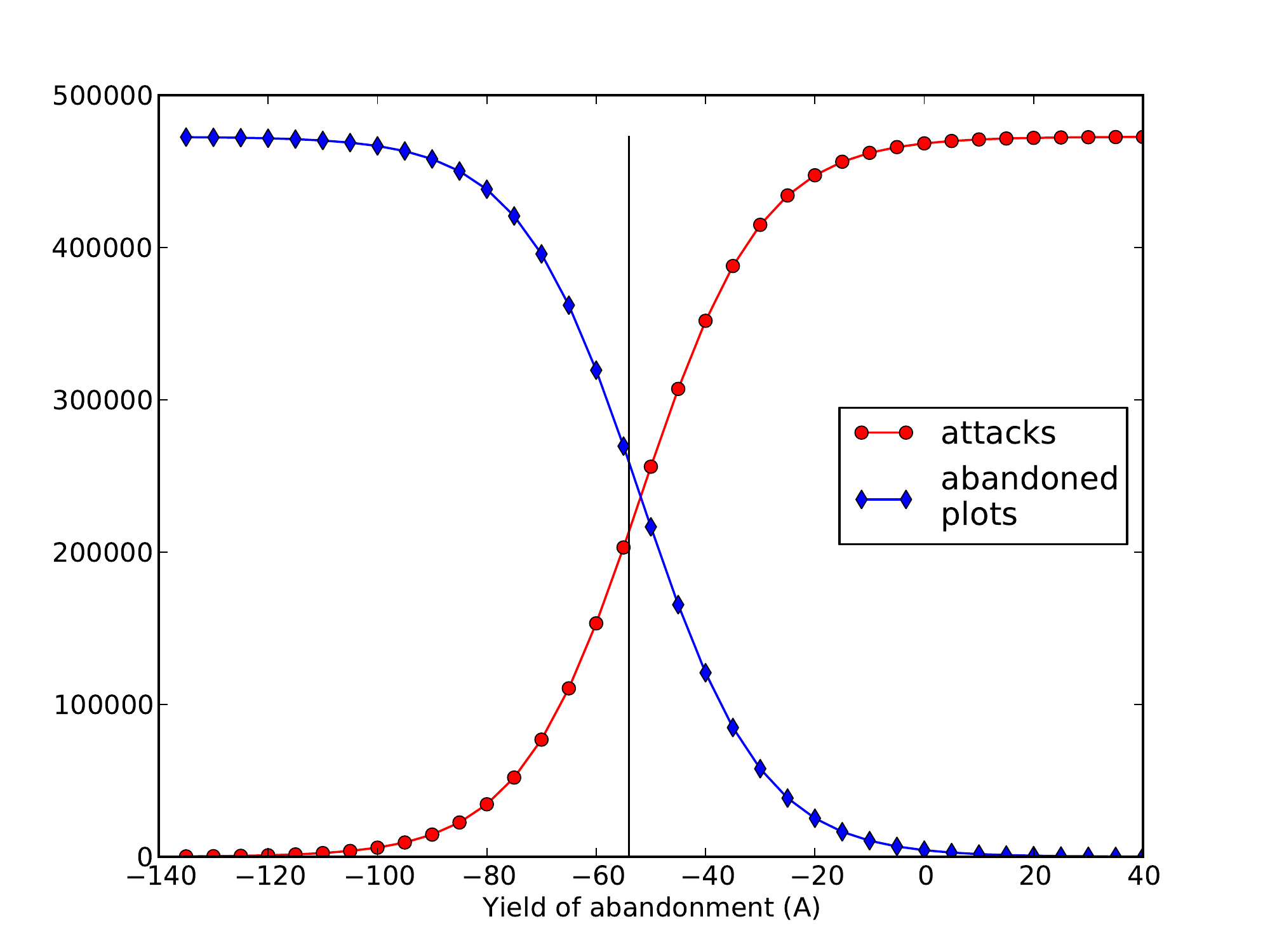}
\par\end{centering}

\caption{The number of attacks as a function of the yield from abandoning.
Negative values make abandoning more competitive and decrease attacks
(left side) while positive values indicate that abandoning is costly
and encourage attacks (right side). The vertical black line indicates
the yield from attacking the US - the most valuable target. The sigmoid
shapes suggest that the effect of deterrence is low until a threshold
is reached, but the threshold must be close to the perceived value
of attacking the US.\label{fig:attacks-vs-A} }

\end{figure}

A number of defensive strategies are founded on deterrence. In terrorism,
deterrence may involve convincing would-be groups or cells that operations
are too risky or that the entire struggle they wage is hopeless. The
model can express such conditions on a global level by varying the
parameter $A$, the perceived yield from abandoning. Raising this
yield is equivalent to raising risks throughout the network. The effect
of $A$ is non-linear, showing a threshold at around $A=-35$ beyond
which attacks decline (Fig.~\ref{fig:attacks-vs-A}.) Unfortunately,
the threshold lies quite high, indeed nearer to the yield from attacking
the US ($-54$) than countries such as France ($-6.8$), implying
that it would be necessary to create a very large deterrence effect
to reduce the number of plots.

If this level of deterrence is somehow achieved, the reduction in
attacks will not occur at once in all countries because cells in some
countries have lower translocation costs than cells in other countries.
As a result, their perceived net benefit and probability of success
are higher. Thus plots originating within the developed countries
such as the G7 and especially home-grown plots will be the last to
experience deterrence because they originate so close to high-value
targets.

\section{Discussion}

The network model draws attention to differences between our past
experience with terrorism and its possible future. Populous regions
like Nigeria and Bangladesh are predicted to produce many plots although
they have not participated significantly in transnational terrorism
yet. If those regions start producing terrorists at a level commensurate
with their size and radicalization, the world will see many more attacks.
Alternatively, it is possible that those regions have characteristics
that hold back violent extremism. If so, future research should identify
those characteristics and suggest policies that maintain and encourage
them.

The model confirms the significant danger from substitution effects.
Perceived successes in reducing the number of attacks against it may
be due to a strategic redirection by terrorist groups that increases
the risk to other countries. In the scenario where the US deters all
attacks by foreign terrorists, many other countries would experience
a large increase in threats. To an extent this has already been seen
in Europe.

The model introduced above has limitations, some introduced for simplicity,
some that are inherent from its foundation as a network model. Perhaps
the most significant shortcoming is the assumption that a terrorist
group's main resource are its human resources. In practice, attacks
also require intelligence gathering, training and materials. Those
resources need to be brought together while maintaining motivation
and secrecy. Nevertheless, the model captures some of the probabilistic
analysis used by transnational groups within the well-developed framework
of network theory.

\section{Conclusions}

The paper introduces a model of transnational terrorist groups that
represents operations as a global activity network. It is possible
to estimate the parameters of the model, and then predict the number
of plots directed at OECD target countries from countries throughout
the world. The model highlights the exceptionally high risk of attacks
against the US. If the US is successful in deterring attacks against
itself without reducing the overall supply of terror then most OECD
countries would see sharp increases in attacks because of a substitution
effect.

The scale of the substitution effect calls for studies into counter-terrorism
policies that are designed to anticipate and sterilize it. Fortunately,
the problem, termed {}``network interdiction'' has already been
studied in the context of other networks \cite{Washburn95,Wood09}.
Those mathematical methods can identify where barriers can be erected
in the transnational terrorist network to produce an increase in the
costs to the terrorists in such a way that they cannot avoid it by
shifting their plots to other countries. Therefore it should be possible
to develop efficient and game-theoretically robust multi-national
defense strategies.

\subsection*{Acknowledgments}

A conversation with Gordon Woo has inspired this work. Matthew Hanson
and Vadas Gintautas suggested useful improvements. Part of this work
was funded by the Department of Energy at the Los Alamos National
Laboratory under contract DE-AC52-06NA25396 through the Laboratory
Directed Research and Development program, and by the Defense Threat
Reduction Agency grant {}``Robust Network Interdiction Under Uncertainty''.
Released as Los Alamos Unclassified Report 10-05689.

\bibliographystyle{amsplain}
\bibliography{terror,interdict,ttn}

\newpage{}

\appendix

\section{Parameter Estimates\label{sec:Parameter-Estimates}}

\begin{flushleft}
\begin{table}[H]
\caption{Information about countries: the interception cost (left) and the
yields from attacks (right). The interception cost ($I_{j}$) is not
known for all OECD states. For this reason, only the countries where
it was available (the left table) were considered as possible targets.
\label{tab:country-tables}}

\end{table}

\par\end{flushleft}

\vspace{-0.3in}

\begin{flushleft}
\begin{minipage}[t]{0.4\columnwidth}%
\begin{center}
\begin{tabular}{|c|c|} \hline Country & Intercept.~Cost $I_j$\tabularnewline \hline \hline New Zealand & 2.3\tabularnewline \hline United Kingdom & 2.1\tabularnewline \hline Czech Republic & 1.6\tabularnewline \hline Hungary & 1.6\tabularnewline \hline United States & 1.5\tabularnewline \hline Slovakia & 1.5\tabularnewline \hline Estonia & 1.5\tabularnewline \hline Portugal & 1.4\tabularnewline \hline Italy & 1.3\tabularnewline \hline Spain & 1.3\tabularnewline \hline Poland & 1.2\tabularnewline \hline Netherlands & 1.2\tabularnewline \hline Israel & 1.1\tabularnewline \hline Belgium & 1.1\tabularnewline \hline Slovenia & 1.0\tabularnewline \hline Germany & 1.0\tabularnewline \hline Canada & 0.9\tabularnewline \hline Austria & 0.9\tabularnewline \hline Iceland & 0.8\tabularnewline \hline Ireland & 0.8\tabularnewline \hline Japan & 0.8\tabularnewline \hline Sweden & 0.7\tabularnewline \hline Finland & 0.6\tabularnewline \hline France & 0.6\tabularnewline \hline South Korea & 0.6\tabularnewline \hline Greece & 0.5\tabularnewline \hline Denmark & 0.3\tabularnewline \hline Luxembourg & 0.2\tabularnewline \hline Norway & 0.2\tabularnewline \hline Australia & 0.0\tabularnewline \hline \end{tabular} 
\par\end{center}%
\end{minipage}\hfill{}%
\begin{minipage}[t]{0.3\columnwidth}%
\begin{center}
\begin{tabular}{|c|c|} \hline Country & Yield $Y_j$\tabularnewline \hline \hline United States & -54.0\tabularnewline \hline Japan & -24.1\tabularnewline \hline Germany & -9.5\tabularnewline \hline United Kingdom & -7.8\tabularnewline \hline France & -6.8\tabularnewline \hline Italy & -5.3\tabularnewline \hline Canada & -3.8\tabularnewline \hline Spain & -3.2\tabularnewline \hline South Korea & -3.1\tabularnewline \hline Australia & -2.2\tabularnewline \hline Netherlands & -1.8\tabularnewline \hline Sweden & -1.2\tabularnewline \hline Belgium & -1.1\tabularnewline \hline Austria & -0.9\tabularnewline \hline Poland & -0.9\tabularnewline \hline Norway & -0.8\tabularnewline \hline Denmark & -0.7\tabularnewline \hline Greece & -0.6\tabularnewline \hline Finland & -0.6\tabularnewline \hline Ireland & -0.5\tabularnewline \hline Portugal & -0.4\tabularnewline \hline Czech Republic & -0.2\tabularnewline \hline New Zealand & -0.2\tabularnewline \hline Hungary & -0.2\tabularnewline \hline Slovakia & -0.0\tabularnewline \hline Luxembourg & -0.0\tabularnewline \hline \end{tabular} 
\par\end{center}%
\end{minipage}
\par\end{flushleft}

\smallskip{}

\begin{table}[H]
\centering{}\caption{Translocation costs $T_{ij}$ (country $i$ to country $j$) for select
country pairs. Rows are countries of departure, columns are the destinations\label{tab:Transit-cost}.
Notice that Japan has relatively large barriers, as estimated by its
abnormally low population of immigrants.}

\end{table}

\hspace{-1.5in}\begin{tabular}{|c|c|c|c|c|c|c|c|c|c|c|} \hline                                   Destination --> & Australia & Canada & France & Germany & Italy & Japan & South Korea & Spain & UK & US\tabularnewline \hline                                                                                                                 \hline                                                                                                                 Afghanistan & 0.0 & 0.0 & 1.9 & 0.1 & 29.5 & 44.2 & 44.2 & 18.1 & 0.3 & 0.1\tabularnewline                             \hline                                                                                                                 Algeria & 0.2 & 0.1 & 0.1 & 9.9 & 15.1 & 32.4 & 32.4 & 9.2 & 6.2 & 1.9\tabularnewline                                  \hline                                                                                                                 Azerbaijan & 0.6 & 0.3 & 10.3 & 4.3 & 50.2 & 71.6 & 71.6 & 9.3 & 5.1 & 0.2\tabularnewline \hline Bangladesh & 0.4 & 0.1 & 8.1 & 3.2 & 1.2 & 12.9 & 8.9 & 5.4 & 0.1 & 0.3\tabularnewline \hline Burkina Faso & 6.3 & 2.0 & 1.1 & 5.4 & 2.5 & 192.0 & 192.0 & 27.8 & 44.1 & 11.2\tabularnewline \hline Chad & 2.8 & 0.6 & 0.7 & 12.2 & 43.9 & 407.2 & 407.2 & 62.4 & 15.3 & 9.3\tabularnewline \hline Cote d'Ivoire & 2.1 & 0.5 & 0.1 & 2.8 & 0.8 & 17.7 & 17.7 & 8.1 & 1.5 & 1.2\tabularnewline \hline Egypt & 0.0 & 0.1 & 1.9 & 1.1 & 2.7 & 12.8 & 12.8 & 13.9 & 1.4 & 0.2\tabularnewline \hline France & 0.0 & 0.1 & 0.0 & 34.6 & 1.9 & 1.8 & 2.7 & 1.3 & 28.2 & 0.2\tabularnewline \hline Guinea & 2.2 & 0.4 & 0.3 & 0.4 & 3.4 & 4.8 & 4.8 & 0.6 & 8.3 & 1.2\tabularnewline \hline India & 0.2 & 0.1 & 4.9 & 3.5 & 5.9 & 64.4 & 153.1 & 9.4 & 0.3 & 0.2\tabularnewline \hline Indonesia & 0.3 & 0.2 & 2.5 & 0.9 & 7.6 & 5.0 & 3.2 & 7.0 & 1.3 & 0.3\tabularnewline \hline Iran & 0.0 & 0.0 & 1.2 & 0.5 & 4.4 & 3.1 & 3.1 & 4.0 & 0.5 & 0.1\tabularnewline \hline Iraq & 0.0 & 0.0 & 3.0 & 0.5 & 13.6 & 98.3 & 98.3 & 5.3 & 0.3 & 0.1\tabularnewline \hline Jordan & 0.0 & 0.0 & 3.5 & 1.4 & 2.9 & 9.2 & 9.2 & 1.2 & 0.8 & 0.0\tabularnewline \hline Kazakhstan & 0.5 & 0.1 & 7.4 & 23.1 & 5.9 & 81.6 & 81.6 & 8.8 & 2.9 & 0.4\tabularnewline \hline Libya & 0.0 & 0.1 & 6.3 & 4.3 & 0.8 & 55.4 & 55.4 & 15.1 & 0.7 & 0.4\tabularnewline \hline Malaysia & 0.0 & 0.0 & 0.8 & 0.1 & 4.1 & 1.8 & 14.8 & 3.7 & 0.0 & 0.1\tabularnewline \hline Morocco & 0.2 & 0.1 & 0.1 & 1.0 & 0.3 & 13.1 & 13.1 & 0.7 & 3.3 & 0.6\tabularnewline \hline Mozambique & 0.7 & 0.4 & 1.8 & 0.8 & 3.8 & 472.8 & 472.8 & 2.0 & 0.4 & 1.7\tabularnewline \hline Nigeria & 0.7 & 0.5 & 14.3 & 1.8 & 3.1 & 8.3 & 8.3 & 4.4 & 0.4 & 0.4\tabularnewline \hline Pakistan & 0.2 & 0.1 & 2.3 & 1.6 & 2.2 & 11.6 & 9.7 & 1.4 & 0.1 & 0.2\tabularnewline \hline Palestine & 0.0 & 0.0 & 2.7 & 1.5 & 15.5 & 1e+200 & 1e+200 & 3.9 & 0.7 & 0.0\tabularnewline \hline Russia & 0.1 & 0.1 & 6.8 & 38.3 & 6.8 & 10.3 & 10.3 & 3.1 & 7.5 & 0.2\tabularnewline \hline Saudi Arabia & 0.2 & 0.1 & 4.4 & 3.7 & 20.7 & 32.4 & 32.4 & 15.6 & 0.6 & 0.2\tabularnewline \hline Senegal & 0.4 & 0.4 & 0.0 & 1.3 & 0.1 & 6.3 & 6.3 & 0.4 & 4.7 & 0.8\tabularnewline \hline Somalia & 0.0 & 0.0 & 0.8 & 0.0 & 0.2 & 463.3 & 463.3 & 6.0 & 0.0 & 0.0\tabularnewline \hline Sudan & 0.1 & 0.1 & 7.9 & 2.8 & 23.6 & 34.5 & 34.5 & 34.5 & 0.8 & 0.5\tabularnewline \hline Tajikistan & 2.3 & 1.5 & 56.7 & 27.3 & 29.8 & 1967.0 & 1967.0 & 80.3 & 12.3 & 0.6\tabularnewline \hline Tanzania & 0.3 & 0.0 & 7.9 & 1.0 & 6.2 & 19.5 & 19.5 & 24.5 & 0.1 & 0.6\tabularnewline \hline Tunisia & 0.2 & 0.1 & 0.1 & 1.5 & 2.4 & 7.8 & 7.8 & 19.2 & 5.2 & 0.8\tabularnewline \hline Turkey & 0.0 & 0.2 & 0.4 & 0.1 & 24.4 & 10.8 & 10.8 & 38.2 & 1.1 & 0.3\tabularnewline \hline UK & 0.0 & 0.0 & 32.2 & 23.7 & 2.9 & 0.7 & 2.7 & 1.3 & 0.0 & 0.1\tabularnewline \hline US & 0.0 & 10.6 & 1.2 & 0.9 & 0.7 & 0.7 & 0.9 & 1.7 & 0.3 & 0.0\tabularnewline \hline \end{tabular} 

\begin{flushleft}
\begin{table}[H]
\caption{The supply of plots for the default weight and change under two alternative
weightings (high commitment and low commitment.) Certain countries
are unusually dependent on the level of support the most radical segment
provides, while others see relatively broad support for violence.
Only the 30 largest sources are shown. For some countries in a particular
region the sensitivity is identical because direct survey data was
not always available. In those countries the radicalization values
($\sigma^{r},\sigma^{s},\sigma^{o}$) were imputed from regional estimates.\label{tab:country-supply}}
 \hspace{-0.8in}

\raggedright{}\begin{tabular}{|c|c|c|c|} \hline Country & Supply $S_i$ & High commitment & Low commitment \tabularnewline \hline \hline Indonesia & 52745.4 & -46.2\% & 24.6\%\tabularnewline \hline Nigeria & 52687.8 & -33.3\% & 17.8\%\tabularnewline \hline India & 49624.7 & -43.1\% & 23.0\%\tabularnewline \hline Bangladesh & 39960.8 & -33.8\% & 18.0\%\tabularnewline \hline Iran & 26723.7 & -30.6\% & 16.3\%\tabularnewline \hline Egypt & 24731.6 & -41.0\% & 21.8\%\tabularnewline \hline Pakistan & 16537.8 & -22.1\% & 11.8\%\tabularnewline \hline Algeria & 12387.6 & -30.6\% & 16.3\%\tabularnewline \hline Iraq & 11021.7 & -30.6\% & 16.3\%\tabularnewline \hline Sudan & 10910.5 & -30.6\% & 16.3\%\tabularnewline \hline Afghanistan & 10168.3 & -30.6\% & 16.3\%\tabularnewline \hline Saudi Arabia & 9037.1 & -30.6\% & 16.3\%\tabularnewline \hline Yemen & 8462.6 & -30.6\% & 16.3\%\tabularnewline \hline Ethiopia & 8278.6 & -39.7\% & 21.2\%\tabularnewline \hline Mali & 8247.4 & -23.2\% & 12.4\%\tabularnewline \hline Uzbekistan & 8161.3 & -43.1\% & 23.0\%\tabularnewline \hline Syria & 7315.4 & -30.6\% & 16.3\%\tabularnewline \hline Morocco & 6878.5 & -26.5\% & 14.1\%\tabularnewline \hline China & 6680.7 & -43.1\% & 23.0\%\tabularnewline \hline Niger & 6497.3 & -32.2\% & 17.2\%\tabularnewline \hline Malaysia & 6466.6 & -47.7\% & 25.4\%\tabularnewline \hline Turkey & 5521.4 & -44.0\% & 23.5\%\tabularnewline \hline Russian Federation & 5081.9 & -43.1\% & 23.0\%\tabularnewline \hline Palestine & 4632.0 & -21.1\% & 11.2\%\tabularnewline \hline Burkina Faso & 4004.9 & -32.2\% & 17.2\%\tabularnewline \hline Tunisia & 3700.5 & -30.6\% & 16.3\%\tabularnewline \hline Senegal & 3668.5 & -40.3\% & 21.5\%\tabularnewline \hline Guinea & 3664.4 & -32.2\% & 17.2\%\tabularnewline \hline Cote d'Ivoire & 3338.1 & -32.2\% & 17.2\%\tabularnewline \hline Somalia & 3258.2 & -30.6\% & 16.3\%\tabularnewline \hline \end{tabular} 
\end{table}
In certain countries, such as Malaysia, Indonesia and Turkey support
for violence is relatively broad. This can be seen from the large
decrease in supply under the high-commitment scenario, ($s_{r},s_{s},s_{o}$)=($0.1,0.2,1.0$),
compared to the default weights ($s_{r},s_{s},s_{o}$)=($0.25,0.50,1.00$).
In regions such as the Palestinian Territories, Pakistan and Morocco
the support is more dependent on the radical minority, as seen from
the relatively small increase under the scenario ($s_{r},s_{s},s_{o}$)=($0.33,0.66,1.00$).
Overall, the $10$ largest sources of plots are not more sensitive
to those parameters than other sources.
\par\end{flushleft}

\section{Computation of Probabilities\label{sub:Computation-of-Probabilities}}

In the framework of the theory of complex networks, attack plots could
be represented as the motion of an adversary through a weighted network
(the plot itself is the adversary we wish to stop.) The adversary
aims to find an attack path or to hide, whichever plan has the lowest
cost. To map such a decision to the framework of activity networks,
connect the {}``attack'' and {}``abandon'' nodes in Fig.~\ref{fig:model-illus}
to a node termed {}``end'' with edges of cost $0$. Thus, an attack
on a country $j$ corresponds to an adversary that starts at country
$i$ and goes through country $j$ and then to the node {}``attack''
and finally to {}``end''. The decision to abandon corresponds to
an adversary that starts at country $i$ then goes to {}``abandon''
and then to {}``end''. The expected number of attacks on a particular
target $t$ can be computed by combining information about path costs
with information about the supply of plots from a particular country
$S_{i}$ and the yield of abandonment $A$. Namely, it is the number
of trips from all sources that arrive to the {}``attack'' node from
target country $t$.

The least-cost path corresponds to the optimal choice by the terrorists,
but they can make mistakes. An attack plan under uncertainty could
be described as a Markov chain \cite[Ch. 3]{Gutfraind09thesis}. The
chain has initial distribution proportional to $S_{i}$, and a transition
probability matrix ${\bf M}$ describing the likelihood of taking
a particular edge on the network. The {}``end'' node is the absorbing
state of the chain. The ${\bf M}$ matrix can be computed using the
least-cost guided evader model described in \cite[Ch. 3]{Gutfraind09thesis}.
Briefly, for each edge $(u,v)$ of the network, the transition probability
through it is given by the formula $M_{uv}\propto\exp(-\lambda\left(c(p_{uv})-c(p_{u*})\right)$,
where $c(p_{u*})$ is the cost of the least-cost path from node $u$
to the end node, and $c(p_{uv})$ is the cost of the path through
edge $(u,v)$: $p_{uv}=(u,v)\cup p_{v*}$. Thus, the model generalizes
the least-cost path model%
\footnote{To compute the distances to the {}``end'' node, $c(p_{uv})$, we
use the Bellman-Ford algorithm because edge weights are negative for
some edges (e.g. yield from attacks). Ref.~\cite[Ch. 3]{Gutfraind09thesis}
uses the faster algorithm of Dijkstra because it treats only the case
of positive weights.%
}. The parameter $\lambda$ was set to $0.1$, in the reported data,
but its value has a smooth effect on the predictions of the model
because of the smoothness of the exponential function. The number
of plots against a target country $t$ is now found by taking the
probability of a trip to that target multiplied by the total number
of plots ($=\sum_{i}S_{i}$).

\end{document}